\documentclass[a4paper,10pt,superscriptaddress,twocolumn,pra]{revtex4-2}

\usepackage{amsmath}            
\usepackage{amssymb}            
\usepackage{amsfonts}           
\usepackage{bm}                 
\usepackage{siunitx}            
\usepackage{graphicx}           
\usepackage{xcolor}             
\usepackage{hyperref}           
\usepackage[version=3]{mhchem}  
\usepackage[mathlines]{lineno}  
\usepackage{epstopdf}           

\newcommand{\dd}{\,\mathrm{d}}

\begin{document}
\title{The Quantum Self-Consistent Harmonic Approximation: A Unified Framework for Quantum Spin Systems}
\author{G. C. Villela}
\email{gabriel.villela@ufv.br}
\affiliation{Departamento de Física, Universidade Federal de Viçosa, 36570-900, Viçosa, Minas Gerais, Brazil}
\author{A. R. Moura}
\email{antoniormoura@ufv.br}
\affiliation{Departamento de Física, Universidade Federal de Viçosa, 36570-900, Viçosa, Minas Gerais, Brazil}
\date{\today}

\begin{abstract}
The Self-Consistent Harmonic Approximation (SCHA) has been utilized to investigate quantum and thermal phase transitions within magnetic models and, 
more recently, in spintronic applications. The SCHA methodology involves utilizing simple harmonic Hamiltonians, which are augmented with 
renormalization parameters that incorporate high-order fluctuations typically overlooked by conventional Linear Spin-Wave (LSW) theories.
Although this approach exhibits reasonable accuracy for models defined by large spin values, its reliability diminishes when applied to quantum 
systems with $S=1/2$. The traditional development of SCHA has incorporated semiclassical assumptions that obscure quantum effects. In this study, 
we introduce a quantum framework for the SCHA that eliminates the need for semiclassical approximations. Our Quantum Self-Consistent Harmonic 
Approximation (QSCHA) utilizes the spin coherent states formalism within a fully quantum formulation. Consequently, we derive a novel 
renormalization parameter that accurately integrates quantum corrections. To assess the efficacy of this new approach, we apply the 
QSCHA to analyze the critical temperature transitions across various well-documented magnetic models. The findings, combined with the simplified 
operational procedure relative to other conventional interacting spin-wave methodologies, suggest that QSCHA is a promising tool for advancing 
research in quantum magnetism and spintronics.
\end{abstract}

\maketitle

\section{Introduction and motivation}
\label{sec.introduction}

Quantum magnetism remains one of the most active areas in condensed matter physics, providing the theoretical framework for 
understanding correlated phenomena such as magnetic ordering, quantum phase transitions, and collective excitations. Additionally,
the investigation of quantum magnetism is a fundamental piece in the development of spintronics \cite{jmmm509.166711,rmp76.323,rmp90.015005}. 
The microscopic description of these effects is commonly formulated in terms of the Heisenberg Hamiltonian, which encapsulates the 
exchange interactions between localized spins in insulating magnetic materials. However, despite its apparent simplicity, this model
exhibits a rich variety of behaviors that challenge both analytical and numerical methods, particularly in low-dimensional or frustrated systems. 
Over the past few decades, significant effort has been devoted to developing approximate schemes capable of capturing quantum and thermal 
fluctuations beyond mean-field treatments. The continuous search for more accurate and efficient methods to describe the quantum 
dynamics of spin systems underscores the importance of exploring new theoretical approaches to the Heisenberg model.

Several theoretical frameworks have been developed to address the complexity of quantum spin systems. Linear and nonlinear 
spin-wave theories, based on bosonic representations such as the Holstein-Primakoff \cite{pr58.1098} 
or Dyson-Maleev \cite{pr102.1217,jetp6.776} representations, provide valuable insights into low-temperature regimes where quantum 
fluctuations are small. Alternatively, the Schwinger-boson formalism \cite{jap67.5734,prb38.316}
extends this treatment by preserving spin rotational symmetry and enabling the study of disordered or frustrated phases. Beyond 
these discrete-spin approaches, field-theoretical descriptions, which include the nonlinear sigma model and path integral formulations, 
offer a continuum perspective suitable for long-wavelength excitations and renormalization-group analysis \cite{auerbach,pires}. 
Despite their success, these methods face limitations in describing intermediate-temperature behavior or strongly anharmonic 
regimes, motivating the development of self-consistent and variational schemes that incorporate quantum and thermal effects on an equal footing.

In recent history, the Self-Consistent Harmonic Approximation (SCHA) has been effectively utilized to assess the critical temperature 
\cite{prb49.9663,prb51.16413,prb54.3019,prb59.6229}, the topological Berezinskii-Kosterlitz-Thouless (BKT) transition 
\cite{prb48.12698,prb49.9663,prb50.9592,ssc100.791,prb53.235,prb54.3019,prb54.6081,prb78.212408,jmmm452.315}, 
and the large-D quantum phase transition \cite{pasma373.387,jpcm20.015208,pasma388.21,pasma388.3779,jmmm357.45} within a diverse range of 
magnetic models. Within the SCHA framework, the Hamiltonian is expressed in terms of a second-order expansion concerning the operators 
$\hat{\varphi}$ and $\hat{S}^z$. The influence of higher-order perturbations is incorporated through renormalization parameters that exhibit 
temperature dependence, which are subsequently resolved via a self-consistent integral equation. Consequently, the SCHA retains the advantages 
inherent to a quadratic Hamiltonian while incorporating corrections from higher-order spin-wave interactions. 
Furthermore, it has been established by Moura and Lopes that the SCHA is fully compatible with the coherent state 
approach \cite{jmmm472.1}. As a result, the SCHA formalism represents a viable option for investigating magnetization precession phenomena 
applied in spintronic processes \cite{prb106.054313,jmmm606.172393}.

Notwithstanding the relative achievements observed, a question concerning the development of the Quantum Self-Consistent Harmonic Approximation
(QSCHA) remains unresolved. Typically, the self-consistent equation for the renormalization parameter is derived within the semiclassical 
approach, where spin operators are replaced by vector fields. Upon deriving the self-consistent equations, we return to the quantum regime by 
substituting classical Gaussian averages with their quantum counterparts. Consequently, the resultant solution is only partially accurate, 
necessitating further corrections to achieve precise quantitative results, depending on the specific model under investigation. In this study, 
we present a comprehensive demonstration of the QSCHA without the necessity for semiclassical 
approximations. The novel self-consistent method is analogous to the conventional approach but incorporates a quantum correction factor that 
cannot be derived from a semiclassical standpoint. We employ the new QSCHA in various scenarios, and the results obtained demonstrate significant 
improvements when compared with data acquired from MC simulations and experimental measurements.

\section{Model description}
\label{sec.model}

To characterize an insulating magnetic material, we employ the anisotropic Heisenberg Hamiltonian, or XXZ model, expressed in terms of dimensionless 
spin operators, as follows:
\begin{align}
    \hat{H}=\pm \frac{J}{2}\sum_{\langle i,j\rangle}(\hat{S}_i^+\hat{S}_j^-+\hat{S}_i^-\hat{S}_j^+ +2\lambda \hat{S}_i^z\hat{S}_j^z),
\end{align}
where $-J$ corresponds to the ferromagnetic (FM) model, whereas $+J$ denotes the antiferromagnetic (AFM) model. In this discussion, 
we shall adhere to the convention whereby the lower signal in $\pm$ (or $\mp$) refers to the FM model, while the upper signal 
denotes the AFM model. The sum is performed exclusively over nearest neighbor interactions in a periodic lattice with lattice spacing $a$. 
The focus of this study is on scenarios involving a small $S^z$ component, thereby justifying the 
use of easy-plane anisotropy characterized by $\lambda<1$. Furthermore, the investigation concerns the thermodynamics of ordered states, which requires spontaneous symmetry breaking 
at temperatures below the critical threshold $T_c$. Consequently, one could designate the x-axis as the spontaneously chosen direction for 
spin alignment in the system, implying that the angle $\varphi$, canonically conjugate to $S^z$, remains small. Analogous consideration is necessary for the 
analysis of easy-axis models. In this context, by identifying the $x$-direction as the easy axis, the formalism can be employed with only minor modifications. 
Nevertheless, this assumption is not essential for the subsequent analysis, and it is sufficient to consider a regime characterized by small, slowly varying angular 
deviations. The inclusion of magnetic field interactions or various alternative anisotropies is feasible, and this process necessitates merely a reassessment 
of the spectral energy.

In the spin formalism, one may attempt to introduce an angle operator $\hat{\varphi}$ that is conjugate to the spin projection operator $\hat{S}^z$, 
in analogy with the canonical commutation relation between position and momentum $[\hat{x}, \hat{p}] = i\hbar$. The intuitive goal is to write a 
similar relation $[\hat{\varphi}_i, \hat{S}_j^z]=i\delta_{ij}$, suggesting that $\hat{\varphi}_i$ represents the angular coordinate of the spin 
on the site $i$. However, such a definition encounters fundamental difficulties; unlike the position-momentum pair, the operator $\hat{S}^z$ 
possesses a discrete and bounded spectrum with eigenvalues $m = -S, -S+1, \dots,S-1, S$. Consequently, it is impossible to define a 
self-adjoint operator $\hat{\varphi}$ that is truly conjugate to $\hat{S}^z$ while maintaining periodicity and proper Hermiticity. 
The angle variable is inherently compact $\varphi \in [0, 2\pi)$, whereas the canonical commutation relation assumes an unbounded conjugate pair.
A similar problem occurs in the investigation of the famous problem of the phase operator \cite{ps48.5,pr6.367}. In the context of spin dynamics, 
Jude and Lewis \cite{pl5.190,nc2.332} showed that the operators $\hat{S}^z$ and the associated angular variable $\hat{\varphi}$ satisfy the 
commutation relation $[\hat{\varphi}_i,\hat{S}_j^z]=i\delta_{ij}[1-2\pi\delta(\varphi_i-\pi)]$, where the angular variable $\hat{\varphi}$ is expressed 
as a 2$\pi$-periodic function of an unbounded angle $\phi$. 

To express the Hamiltonian utilizing canonically conjugate operators that satisfy the commutation relationship $[\hat{\varphi}_i,\hat{S}_j^z]=i\delta_{ij}$, 
we employ the Villain representation \cite{jp35.27}.This allows us to write $\hat{S}_i^+=e^{i \hat{\varphi}_i}\sqrt{\tilde{S}^2-\hat{S}_i^z(\hat{S}_i^z+1)}$ 
and $\hat{S}_i^-=(\hat{S}_i^+)^\dagger$, wherein $\tilde{S}^2=S(S+1)$. It can be demonstrated through a straightforward procedure that
$[e^{\pm i\hat{\varphi}_i},\hat{S}_j^z]=\mp e^{\pm i\hat{\varphi}_i}\delta_{ij}$, and the Villain representation effectively defines a fulfillment 
representation of the spin operators, maintaining the integrity of all spin commutation relations. Provided that $\hat{\varphi}$ and $\hat{S}_i^z$
define small fluctuations around the ordered state, we expand the Hamiltonian up to second-order contributions, resulting in 
$\hat{H}\approx E_0+\hat{H}_1+\hat{H}_2$, where $E_0$ is the unimportant ground-state energy,
\begin{align}
    \hat{H}_1=\mp zJ\sum_i\hat{S}_i^z,
\end{align}
and
\begin{align}
    \hat{H}_2=J\sum_{\langle i,j\rangle}\left[\frac{\tilde{S}^2}{2}(\hat{\varphi}_i-\hat{\varphi}_j)^2+\hat{S}_i^z(\hat{S}_i^z\pm\lambda\hat{S}_j^z)\right].
\end{align}
For the AFM model, we apply a rotation of $\pi$ radians about the z-axis prior to performing the series expansion. As a result, 
we obtain the term $+\lambda\hat{S}_i^z\hat{S}_j^z$, in contrast to the term $-\lambda\hat{S}_i^z\hat{S}_j^z$, which is indicative 
of the FM scenario. It is noteworthy that the operator $\hat{H}_1$ commutes with the quadratic Hamiltonian, thereby establishing a 
conserved quantity for dynamical evolution. Consequently, in subsequent analyzes, we shall restrict our focus exclusively to the 
quadratic term, utilizing it as the model representation.

The quadratic Hamiltonian can be regarded as analogous to a linear spin-wave expansion and serves as a plausible model in the asymptotic 
regime at very low-temperatures. Nevertheless, to incorporate higher-order contributions, we introduce a renormalization parameter 
into the angular expansion by substituting $\hat{\varphi}$ with $\sqrt{\rho}\hat{\varphi}$. The inclusion of a renormalization parameter 
to adjust the square root expansion of $\hat{S}^z$ can be conceptually considered; however, it is not a viable approach. The intrinsic 
oscillatory characteristics of the angular operator are crucial for facilitating the implementation of angle renormalization. 
Additionally, the determination of the parameter $\rho$ constitutes a principal objective of the QSCHA, which will be elaborated upon in the 
next section. Performing the Fourier transform, we obtain the harmonic Hamiltonian expressed as
\begin{align}
    \hat{H}_0=\frac{1}{2}\sum_q(h_q^\varphi\hat{\varphi}_q^\dagger\hat{\varphi}_q+h_q^z\hat{S}_q^{z\dagger}\hat{S}_q^z),
\end{align}
where the coefficients are defined as $h_q^\varphi=2zJ\tilde{S}^2\rho(1-\gamma_q)$ and $h_q^z=2zJ(1\pm\lambda\gamma_q)$. The structure factor 
$\gamma_q=z^{-1}\sum_\eta e^{i\bm{q}\cdot\bm{\eta}}$ is determined by the $z$ nearest-neighbor spins located at $\bm{\eta}$ positions.

The diagonal Hamiltonian is obtained by defining bosonic operators that are responsible for the creation and annihilation of magnons 
via the following relations
\begin{align}
    \hat{\varphi}_q&=\frac{1}{\sqrt{2}}\left(\frac{h_q^z}{h_q^\varphi}\right)^{1/4}(a_{-q}^\dagger+a_q)\\
    \hat{S}_q^z&=\frac{i}{\sqrt{2}}\left(\frac{h_q^\varphi}{h_q^z}\right)^{1/4}(a_{-q}^\dagger-a_q).
\end{align}
It is straightforward to verify that $[a_q,a_{q^\prime}^\dagger]=\delta_{qq^\prime}$, provided that 
$[\hat{\varphi}_q,\hat{S}_{q^\prime}^z]=i\delta_{qq^\prime}$. Then, in terms of the magnon operators, we obtain
\begin{align}
\label{eq.H0}
    \hat{H}_0=\sum_q \epsilon_q\left(a_q^\dagger a_q+\frac{1}{2}\right),
\end{align}
where $\epsilon_q=\hbar\omega_q=2zJ\tilde{S}\sqrt{\rho(1-\gamma_q)(1\pm\lambda\gamma_q})$ denotes the spectrum energy.
In the limit of long wavelengths, it is observed that $\gamma_q\approx 1-q^2/z$ results in a linear energy spectrum $\epsilon=pc$,
as expected from the planar model. Here, $p=\hbar q$ represents the momentum, and $c=(2aJ\tilde{S}/\hbar)\sqrt{z\rho(1\pm\lambda)}$ defines 
the velocity of the spin wave.

The identical outcome is derived from the usual linear Holstein-Primakoff formalism. In this context, by selecting the x-axis as the 
direction of quantization, the spin operators are denoted as $\hat{S}_i^x=S-b_i^\dagger b_i$, $\hat{S}_i^y\approx \sqrt{S/2}(b_i^\dagger+b_i)$, 
and $\hat{S}_i^z\approx i\sqrt{S/2}(b_i^\dagger -b_i)$, leading to the quadratic Hamiltonian
\begin{align}
    \hat{H}_{HP}&=\frac{zJS}{2}\sum_q\left\{[2-(1\mp\lambda)\gamma_q](b_q^\dagger b_q+b_{-q}b_{-q}^\dagger)+\right.\nonumber\\
    &+\left.(-1\mp\lambda)\gamma_q (b_q^\dagger b_{-q}^\dagger+b_{-q}b_q)\right\}.
\end{align}
The Hamiltonian is diagonalized by introducing new bosonic operators through the Bogoliubov transformation $b_q=\cosh\theta_q a_q-\sinh\theta_q a_{-q}^\dagger$, where 
$\tanh2\theta_q=(1\pm\lambda)\gamma_q/[(1\mp\lambda)\gamma_q-2]$ determines the angle required to nullify the off-diagonal terms. 
Consequently, the diagonal HP Hamiltonian is expressed as $\hat{H}_{HP}=\sum_q\epsilon_q(a_q^\dagger a_q+1/2)$, where $\epsilon_q=zJS\sqrt{(1-\gamma_q)(1\pm\lambda\gamma_q)}$ 
represents the identical energy spectrum obtained from the QSCHA when $\rho$ is set to unity and $S$ replaces $\tilde{S}$. In the vicinity of 
the critical temperature, magnon interactions must be considered by incorporating quartic or higher-order terms into the theoretical analysis. 
This integration, however, introduces additional complexity to the analytical process.

Once the Hamiltonian is mapped onto an effective harmonic model, the thermodynamics can be derived from its partition function, 
which takes the generic form associated with a collection of independent quantum harmonic oscillators. The total partition function 
reads:  
\begin{align}
Z_0 = \prod_{q} \left[2\sinh\left(\frac{\beta\epsilon_q}{2}\right)\right]^{-1}.    
\end{align}
From this expression, all thermodynamic quantities can be consistently obtained. The Helmholtz free energy follows as  
$F_0= -k_BT \ln Z_0 = k_{B}T \sum_q \ln\left[2\sinh\left(\beta \epsilon_q/2\right)\right]$, from which one derives the 
internal energy, entropy, and specific heat through standard thermodynamic relations. Within the QSHA framework, the dependence 
of the energies $\epsilon_q$ on the self-consistently determined renormalization parameter ensures that the partition function 
accurately represents the renormalized dynamics of the interacting spin system.

\section{The renormalization parameter}
\label{sec.rho}

In the preceding section, we introduced a renormalization parameter within the context of angle expansion to account for 
the omission of higher-order terms in the series expansion of $\hat{H}$. In this section, we undertake a comprehensive examination of 
the incorporated parameter. In order to determine the $\rho$ equation, we look to the Gibbs-Bogoliubov inequality \cite{ajp38.858}, which 
establishes a variational principle for estimating the upper limit of the free energy $F$ associated with a general 
Hamiltonian $\hat{H}$. It states that $F$ is bounded from above according to $F \leq F_0-\langle\hat{H}_0 \rangle_0+\langle\hat{H}\rangle_0$,
where $\hat{H}_0$ is a suitably chosen trial Hamiltonian for which all thermodynamic averages can be exactly evaluated, and 
$\langle \cdots \rangle_0$ denotes expectation values taken over the ensemble generated by $\hat{H}_0$. With respect to the QSHA, one adopts the 
harmonic Hamiltonian $\hat{H}_0$ endowed with the variational parameter $\rho$ that governs the effective strength of the harmonic fluctuations. 
The optimal value of this parameter is determined by minimizing the function $\Gamma(\rho)=F_0-\langle \hat{H}_0\rangle_0+\langle\hat{H}\rangle_0$,
thereby ensuring that the approximate free energy satisfies the Gibbs–Bogoliubov bound as closely as possible and self-consistently 
incorporates quantum fluctuation effects. From quantum thermodynamics, we achieve
\begin{align}
    \label{eq.F0H0}
    F_0-\langle\hat{H}_0\rangle_0=\sum_q\left[k_B T\ln(1-e^{-\beta\epsilon_q})-\epsilon_qn_q\right],
\end{align}
where $n_q=(e^{\beta\epsilon_q}-1)^{-1}$ represents the Bose-Einstein distribution. 

In the classical approach, when the condition $\epsilon_q\ll k_B T$ is satisfied, the expression $F_0-\langle\hat{H}_0\rangle_0$ can be 
approximated by $\sum_q k_B T(\ln \beta\epsilon_q-1)$. The evaluation of this expression is performed by averaging the Hamiltonian, 
employing $e^{-\beta H_0}$ as a weighting function. Consequently, the function $\Gamma(\rho)$ can be readily determined, 
and the condition $\dd\Gamma/\dd\rho=0$ yields the well-established self-consistent equation given by
\begin{align}
    \rho=\left(1-\frac{\langle(S^z)^2\rangle_0}{S^2} \right)e^{-\langle\Delta\varphi^2\rangle_0/2},
\end{align}
where the averages are determined through simple Gaussian integrals. In this particular scenario, one can determine the mean values 
to deduce the simplified equation $\rho=(1-I t)e^{-t/\rho}$, where the parameter $I$ is expressed as $I=\sum_q(1\pm\lambda\gamma_q)^{-1}/N$.
The reduced temperature is defined by the relation $t=k_BT/(2zJ)$. At the phase transition temperature, $\rho$ experiences a discontinuous 
transition to zero, and the derivative $\dd t/\dd\rho|_{t=t_c}=0$ holds true, yielding the critical temperature $t_c=(I+e)^{-1}$. For a 
classical vector spin model, $t_c$ serves as an excellent approximation for the transition temperature. For example, the classical SCHA yields a critical 
temperature $T_c=4.40 J/k_B$. for the classical XY model and $T_c=2.83 J/k_B$ for the classical Heisenberg model on the simple cubic (SC) lattice. 
In contrast, Monte Carlo (MC) simulations determine $T_c=4.41 J/k_B$ for the XY model and $T_c=2.87 J/k_B$ for the Heisenberg model \cite{pasma201.581}. 
Conversely, in the case of quantum models, mainly for the spin $S=1/2$, a more meticulous analysis is required. Upon achieving the self-consistent equation, the quantum 
version has been derived by replacing the Gaussian averages with quantum statistical averages. However, the quantum framework introduces greater complexities that cannot be 
derived from the semiclassical extension.

\subsection{Spin coherent states formalism}
To properly evaluate the average $\langle\hat{H}\rangle_0$, we employ the spin coherent states formalism, which provides a powerful representation 
of quantum spin systems, establishing a bridge between the discrete spin algebra and the continuous vector fields \cite{perelomov}. A spin coherent state 
$|\theta, \varphi\rangle\equiv|\Omega\rangle$ is obtained by rotating the fully polarized state $|S, S\rangle$ along the direction defined by 
the polar and azimuthal angles $(\theta, \phi)$, namely, $|\Omega\rangle = e^{-i\varphi \hat{S}_z} e^{-i\theta \hat{S}_y} |S, S\rangle$.
This construction ensures that $|\Omega\rangle$ is an eigenstate of the spin projection operator $\hat{\mathbf{S}} \cdot \mathbf{\Omega}$ 
with maximal eigenvalue $S$, where $\mathbf{\Omega} = (\sin\theta\cos\varphi,\, \sin\theta\sin\varphi,\, \cos\theta)$ is a unit vector on the sphere $S^2$. 
The family of states $\{|\Omega\rangle\}$ constitutes an overcomplete basis of the Hilbert space and satisfies the resolution of the identity:
\begin{align}
\hat{I}=\frac{2S+1}{4\pi} \int \dd\Omega\, |\Omega\rangle \langle \Omega|,    
\end{align}
where $\dd\Omega = \sin\theta\, \dd\theta\, \dd\varphi$ represents the solid angle element. This property allows one to represent traces, expectation values, 
and thermodynamic quantities as integrals over the continuous variables $(\theta, \varphi)$. The spin operators themselves take the simple 
form $\langle \Omega | \hat{\mathbf{S}} | \Omega \rangle = S\, \mathbf{\Omega}$, showing that the coherent state behaves as a spin vector of 
magnitude $S$ pointing along $\mathbf{\Omega}$. These properties make the spin coherent states an essential tool for constructing path integrals in spin 
space. 

Adopting the imaginary time formalism, the spin coherent states establish the partition function as the path spin integral
$Z=\int\mathcal{D}\Omega e^{-\mathcal{A}/\hbar}$, where the integral measurement is expressed as $\mathcal{D}\Omega=\prod_i \dd\Omega_i$. 
Herein, the action is defined as
\begin{align}
    \mathcal{A}=\int_0^{\beta\hbar}\left[i\hbar\sum_iS_i^z\dot{\varphi}_i -H(\tau)\right]\dd\tau,
\end{align}
where $H(\tau)$ denotes the expectation value $H(\tau)=\langle\Omega|\hat{H}|\Omega\rangle$. Given that the Hamiltonian $\hat{H}$ is 
linear with respect to the spin operators, $H(\tau)$ can be derived by substituting the operator $\hat{S}_i^\alpha$ with the classical field 
$S_i^\alpha$. It is noteworthy that non-linear spin operators do not yield expectation values that equate to the analogous classical spin values. 
Nonetheless, the discrepancy diminishes with increasing spin magnitude $S$, allowing the application of this methodology, at least qualitatively, 
in scenarios involving single-ion anisotropies, for instance. Following the same argument, spin-spin correlations for different sites can be 
expressed as
\begin{align}
    \label{eq.correlation}
    \langle\hat{S}_i^\alpha(\tau)\hat{S}_j^\alpha(\tau^\prime)\rangle=e^{\beta F}\int\mathcal{D}\Omega S_i^\alpha(\tau)S_i^\alpha(\tau^\prime)e^{-\mathcal{A}/\hbar},
\end{align}
where $F=-k_B T\ln Z$, and $\mathbf{S}_i=S\mathbf{\Omega}_i$.

In order to derive a quadratic model that is consistent with the preceding quantum results, we implement a transformation in the integration 
fields characterized by the substitution $\mathbf{\Omega}_i\to(\tilde{S}/S)\mathbf{\Omega}_i$, and 
$\mathcal{D}\Omega\to J_{\Omega\Omega^\prime}\mathcal{D}\Omega^\prime$, where $J_{\Omega\Omega^\prime}$ denotes the Jacobian determinant.
In terms of the new spin fields, the classical Hamiltonian is written as
\begin{align}
    H=-J\sum_{\langle i,j\rangle}\left(S_i^xS_j^x+S_i^yS_j^y\pm\lambda S_i^zS_j^z\right),
\end{align}
where, henceforth, the spin field is defined by $\mathbf{S}_i=\tilde{S}\mathbf{\Omega_i}$. By employing a spin field oriented along the x-axis, 
it is appropriate to express the Hamiltonian as $H=H_2+\varepsilon H^\prime$, where we neglect the constant ground-state energy. The term $H_2(\varphi,S^z)$ 
represents a quadratic Hamiltonian in the variables $\varphi_i$ and $S_i^z$, while $H^\prime$ denotes the higher-order contributions for which 
$\varepsilon\ll1$. Therefore, the partition function is written as $Z=\int\mathcal{D}\Omega\exp[-(\mathcal{A}_0+\mathcal{A^\prime)/\hbar}]$, where
\begin{align}
    \label{eq.action0}
    \mathcal{A}_0=\int_0^{\beta\hbar}\left(i\hbar\sum_iS_i^z\dot{\varphi}_i-H_2\right)\dd\tau,
\end{align}
and $\mathcal{A}^\prime=\int_0^{\beta\hbar} H^\prime \dd\tau$. To the first-order approximation, we derive $Z=Z_0(1-\varepsilon\langle\mathcal{A}^\prime\rangle_0/\hbar)+\mathcal{O}(\varepsilon^2)$,
where the non-interacting spin field average is defined by  $\langle \Psi\rangle_0=Z_0^{-1}\int\mathcal{D}\Omega \Psi e^{-\mathcal{A}_0/\hbar}$.
Consequently, for equal times, the spin-spin correlation function, given by Eq. (\ref{eq.correlation}), provides the expected value
\begin{align}
    \langle\hat{H}\rangle&\approx \frac{1}{Z_0}\left(1+\frac{\varepsilon}{\hbar}\langle\mathcal{A}^\prime\rangle_0 \right)\int\mathcal{D}\Omega e^{-\mathcal{A}_0/\hbar}\left(1+\frac{\varepsilon}{\hbar}\mathcal{A}^\prime \right)H\nonumber\\
    &=\langle H\rangle_0+\frac{\varepsilon}{\hbar}\left(\langle\mathcal{A}^\prime\rangle_0\langle H\rangle_0-\langle\mathcal{A}^\prime H\rangle_0 \right)+\mathcal{O}(\varepsilon^2)
\end{align}
In the subsequent procedures, contributions exceeding the order of $\varepsilon$ are neglected. To account for this, the renormalization 
parameter is incorporated into the angle series expansion, yielding the harmonic field Hamiltonian
\begin{align}
    H_0=J\sum_{\langle i,j\rangle}\left[\frac{1}{2}\tilde{S}^2\rho\Delta\varphi_{ij}^2+S_i^zS_i^z\pm \lambda S_i^z S_j ^z\right],
\end{align}
which replaces $H_2$. The dynamics are derived from the Euler-Lagrange equation, which yields a coupled system of ODEs represented by $\dot{\varphi}_q=h_q^z S_q^z$ and 
$\dot{S}_q^z=-h_q^\varphi\varphi_q$. The solution is characterized by oscillatory fields, where the frequency is given by $\omega_q=\epsilon_q/\hbar$. The same result is
obtained by promoting the angle and $S^z$ fields to canonically conjugate operators. Furthermore, the aforementioned Hamiltonian provides the expected energy spectrum obtained 
from the previous quantum model.

\subsection{Evaluation of $\langle \hat{H}\rangle_0$}
The preceding result provides a justification for considering $\langle\hat{H}\rangle_0$ as given by the field average 
$\langle H\rangle_0$. The planar Hamiltonian contribution involves the term $\zeta_{ij}\cos\Delta\varphi_{ij}$, where $\zeta_{ij}=\zeta(S_i^z,S_j^z)=\sqrt{\tilde{S}^2-(S_i^z) ^2}\sqrt{\tilde{S}^2-(S_j^z) ^2}$.
The average is then expressed as:
\begin{align}
    \label{eq.zetacos}
    \langle\zeta_{ij}\cos\Delta\varphi_{ij}\rangle_0=\textrm{Re}\left[ \frac{1}{Z_0}\int\mathcal{D}\Omega\zeta_{ij}e^{-\mathcal{A}_0/\hbar-i\Delta\varphi_{ij}}\right].
\end{align}
Based on the argument of the exponential, we define the nonlocal action $\mathcal{A}_{ij}=\mathcal{A}_0+i\hbar\Delta\varphi_{ij}$. After performing the Fourier transform,
we obtain
\begin{align}
    \mathcal{A}_{ij}=&\frac{1}{2\beta}\sum_{q,\omega_n}[(i\bar{\Delta}_q^{ij}-\omega_n\bar{S}_{qn}^z)\varphi_{qn}+(i\Delta_q^{ij}+\omega_n S_{qn}^z)\bar\varphi_{qn}+\nonumber\\
    &\left.+\frac{h_q^\varphi}{\hbar}\bar{\varphi}_{qn}\varphi_{qn}+\frac{h_q^z}{\hbar}\bar{S}_{qn}^zS_{qn}^z\right],
\end{align}
where
\begin{align}
    \Delta_q^{ij}=\frac{e^{-i\mathbf{q}\cdot\mathbf{r}_i}-e^{-i\mathbf{q}\cdot\mathbf{r}_j}}{N^{1/2}}
\end{align}
serves as a mechanism for establishing interactions between adjacent sites. The action is substantially simplified by the elimination of mixing terms of 
the type $\bar{\varphi}_{qn}S_{qn}^z$, which is achieved by expressing $\varphi_{qn}=\phi_{qn}+\delta\varphi_{qn}$, where $\phi_{qn}$ signifies the minimum 
of $\mathcal{A}_{ij}$. Then, the condition $\partial\mathcal{A}_{ij}/\partial\varphi_{qn}|_{\varphi=\phi}=0$ yields
\begin{align}
    \phi_{qn}=-\frac{\hbar}{h_q^\varphi}(i\Delta_q^{ij}+\omega_n S_{qn}^z),
\end{align}
with a similar outcome for the independent variable $\bar{\phi}_{qn}$. Furthermore, defining the deviation in $S_{qn}^z$ as
\begin{align}
    \label{eq.deltaSz}
    \delta S_{qn}^z=S_{qn}^z+\frac{i\omega_n}{\omega_q^2+\omega_n^2}\Delta_q^{ij},
\end{align}
allow us to separate the action into distinct components:
\begin{align}
    \mathcal{A}_{ij}=\mathcal{A}_0^\varphi+\mathcal{A}_0^z+\hbar\Xi_{ij},
\end{align}
where we identify the independent actions
\begin{align}
    \mathcal{A}_0^\varphi=\frac{1}{2}\sum_{q,\omega_n} \frac{h_q^\varphi}{\beta\hbar}\delta\bar{\varphi}_{qn}\delta\varphi_{qn},
\end{align}
and
\begin{align}
    \mathcal{A}_0^z=\frac{1}{2}\sum_{q,\omega_n} \frac{\hbar(\omega_q^2+\omega_n^2)}{\beta h_q^\varphi}\delta\bar{S}_{qn}^z\delta S_{qn}^z.
\end{align}    
Additionally, we establish the angle
\begin{align}
    \Xi_{ij}=\frac{1}{2}\sum_{q,\omega_n} \frac{h_q^z\bar{\Delta}_q^{ij}\Delta_q^{ij}}{\beta\hbar^2(\omega_q^2+\omega_n^2)},
\end{align}
which plays an important role in the subsequent thermodynamic analysis.
It is worth noting that the expression $\bar{\Delta}_q^{ij}\Delta_q^{ij}=|\Delta_q^{ij}|^2$ is equivalent to $2(1-\cos\Delta \mathbf{q}\cdot\mathbf{\eta})/N$, 
where $\mathbf{\eta}$ represents the separation between neighboring sites. By averaging over the nearest neighbor interactions, this expression can be written as 
$|\Delta_q^{ij}|^2=2(1-\gamma_q)/N$. Moreover, the sum over the Matsubara frequencies $\omega_n$ yields 
$\langle\bar{\varphi}_q\varphi_q\rangle_0$, as explained in Appendix (\ref{appendix}). Following this, we adopt the replacement $\Xi_{ij}\to\Xi$, where
\begin{align}
    \Xi=\frac{1}{N}\sum_q(1-\gamma_q)\langle\bar{\varphi}_q\varphi_q\rangle_0=\sum_{\langle i,j\rangle}\frac{\langle(\varphi_i-\varphi_j)^2\rangle_0}{2Nz}.
\end{align}

The partition function derived from $\mathcal{A}_0^\varphi$ and $\mathcal{A}_0^z$ coincides with that obtained from the non-interacting model, as elucidated in Appendix (\ref{appendix}).
The expression for $Z_0$ is then given by $Z_0=Z_0^\varphi Z_0^z=\int\mathcal{D}\Omega \exp[-(\mathcal{A}_0^\varphi+\mathcal{A}_0^z)/\hbar]$. 
Returning to Eq. (\ref{eq.zetacos}), we obtain:
\begin{align}
    \langle \zeta_{ij}\cos\Delta\varphi_{ij}\rangle_0=\langle \zeta_{ij}\rangle_0 \exp\left[-\frac{1}{2}\langle\Delta\varphi^2\rangle_0\right],
\end{align}
wherein the mean value $\langle\zeta_{ij}\rangle_0$ solely involves integration over the variable $S^z$. In contrast to the angular component, 
which can be integrated precisely because of its sinusoidal characteristics, the $S^z$ integral cannot be evaluated exactly, necessitating the employ of approximations. 
By utilizing the isotropic properties and the weak spin-wave interaction, it is assumed that 
$\langle \zeta_{ij}\rangle_0\approx \tilde{S}^2-\langle (S^z)^2\rangle_0$, while a spatiotemporal average yields
\begin{align}
\langle (S^z)^2\rangle_0=\frac{1}{N(\beta\hbar)^2}\sum_q\sum_{\omega_n}\langle \bar{S}_{qn}^zS_{qn}^z\rangle_0.    
\end{align}
Utilizing Eq. (\ref{eq.deltaSz}), we derive:
\begin{align}
    \langle (S^z)^2\rangle_0&=\sum_q\sum_{\omega_n}\left[\frac{\langle \delta\bar{S}_{qn}^z\delta S_{qn}^z\rangle_0}{N(\beta\hbar)^2}+\frac{2(1-\gamma_q)}{(N\beta\hbar)^2}\frac{\omega_n^2}{(\omega_n^2+\omega_q^2)^2} \right]\nonumber\\
    &=\frac{1}{N}\sum_q\frac{h_q^\varphi}{2\epsilon_q}\coth\left(\frac{\beta\epsilon_q}{2}\right),    
\end{align}
noting that the term proportional to $N^{-2}$ becomes negligible in the thermodynamic limit. By following an analogous methodology, it is determined for nearest-neighbors ($i\neq j$) that
\begin{align}
\langle S_i^zS_j^z\rangle_0=\frac{1}{N}\sum_q\frac{\gamma_q h_q^\varphi}{2\epsilon_q}\coth\left(\frac{\beta\epsilon_q}{2}\right).    
\end{align}
In conclusion, by aggregating the derived averages, we find:
\begin{align}
\langle \hat{H}\rangle_0&\approx  zJ\sum_q\frac{h_q^\varphi(e^{-\Xi}\pm\lambda\gamma_q)}{2\epsilon_q}\coth\left(\frac{\beta\epsilon_q}{2}\right)-\nonumber\\
&-zNJ\tilde{S}^2.    
\end{align}

\subsection{The self-consistent equation}
Upon computing all necessary averages, the self-consistent equation for $\rho$ is derived from the condition $\dd\Gamma/\dd\rho=0$. Utilizing Eq. (\ref{eq.F0H0}), we derive
\begin{align}
    \frac{\dd}{\dd\rho}(F_0-\langle\hat{H}_0\rangle_0)=\frac{1}{4\rho}\sum_q(v_q-u_q),
\end{align}
where, for convenience, we define the functions
\begin{align}
    \label{eq.u}
    u_q=\epsilon_q\frac{\sinh\beta\epsilon_q-\beta\epsilon_q}{[2\sinh(\beta\epsilon_q/2)]^2},
\end{align}
and
\begin{align}
    \label{eq.v}
    v_q=\epsilon_q\frac{\sinh\beta\epsilon_q+\beta\epsilon_q}{[2\sinh(\beta\epsilon_q/2)]^2}.
\end{align}
The derivative of $\Xi$ is expressed as
\begin{equation}
    \frac{\dd\Xi}{\dd\rho}=-\frac{1}{4zJ\tilde{S}^2\rho^2}\frac{1}{N}\sum_qv_q,
\end{equation}
while the derivative of $\langle(S^z)^2\rangle_0$ is given by
\begin{align}
    \frac{\dd}{\dd\rho}\langle (S_i^z)^2\rangle_0=\frac{1}{N}\sum_q \frac{u_q}{2\rho h_q^z},
\end{align}
with an analogous outcome for $\langle S_i^zS_j^z\rangle_0$. These results culminate in
\begin{align}
    \frac{\dd}{\dd\rho}\langle\hat{H}\rangle_0&=-\frac{1}{4\rho^2}\left[\left(1-\frac{\langle(S^z)^2\rangle_0}{\tilde{S}^2} \right)e^{-\Xi}\sum_qv_q +\right.\nonumber\\
    &\left.+\rho\sum_q\frac{-e^{-\Xi}\pm\lambda\gamma_q}{1\pm\lambda\gamma_q}u_q\right].
\end{align}
Following a straightforward procedure, we derive the self-consistent equation
\begin{align}
    \label{eq.quantumrho}
    \rho(T)=\Lambda(T)\left(1-\frac{\langle(S^z)^2\rangle_0}{\tilde{S}^2} \right)e^{-\Xi},
\end{align}
where
\begin{align}
    \Lambda(T)=\left[\sum_q\left(v_q+\frac{e^{-\Xi}-1}{1\pm\lambda\gamma_q}u_q \right) \right]^{-1}\sum_qv_q.
\end{align}
It is important to observe that, notwithstanding the $\Lambda(T)$ equation, the self-consistent equation derived 
is identical to those obtained from the traditional SCHA.The function $\Lambda(T)$ serves as a quantum 
correction factor, attaining a value of unity within the semiclassical limit. Indeed, in the limit where $\epsilon_q\ll k_B T$, 
it is observed that $v_q \to 1$ and $u_q \to 0$, resulting in the self-consistent equation converging to those derived from 
the conventional SCHA. Eq. (\ref{eq.quantumrho}) constitutes the principal finding of this study. It represents a significant 
improvement in the accuracy of data obtained from the QSCHA, particularly concerning models involving spin $S=1/2$. 
In the next section, we employ the QSCHA methodology to determine the thermodynamic properties across different scenarios.

\section{Results}
\label{sec.results}
Given the quadratic nature of the Hamiltonian, the thermodynamic data can be readily derived from quantum statistical mechanics. 
The thermal and quantum corrections are incorporated via the renormalization parameter specified by Eq. (\ref{eq.quantumrho}), 
which necessitates a numerical solution through a self-consistent iterative method. The convergence of the integral equation is 
rapid, and the temperature dependency of $\rho$ is easily determined.
\begin{figure}[ht]
\centering 
\includegraphics[width=\linewidth]{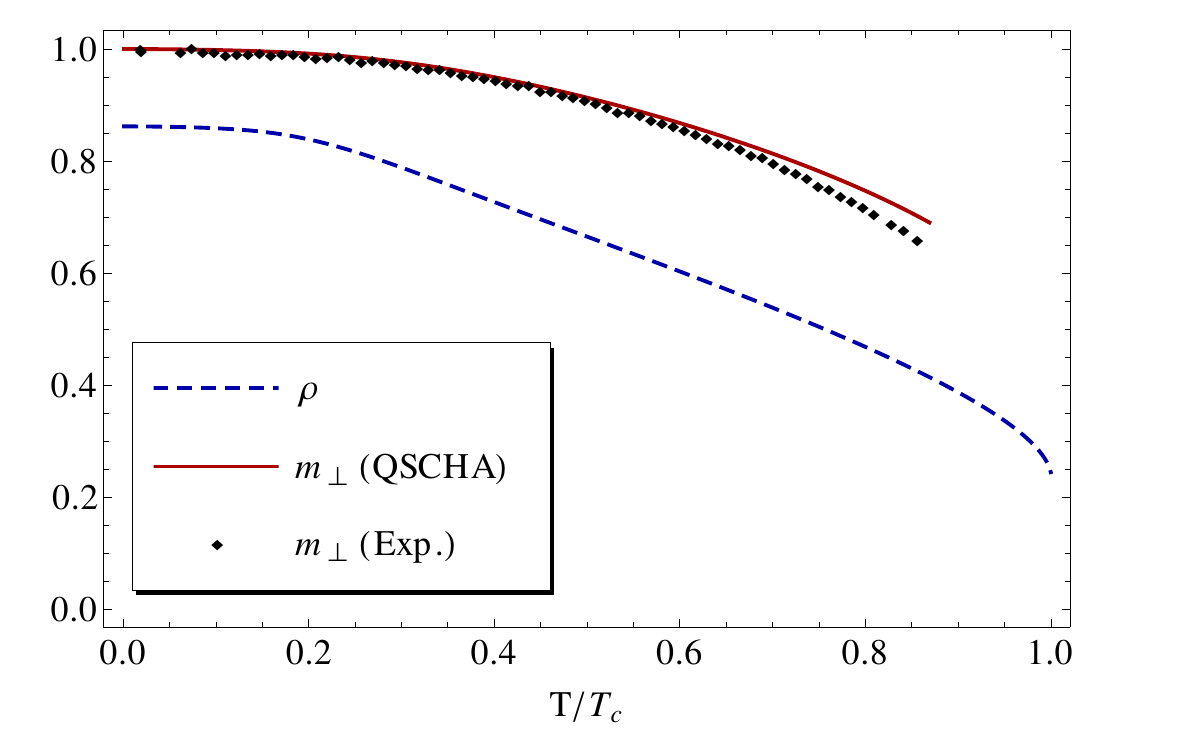}
\caption{The renormalization parameter demonstrates a decreasing trend with the elevation of temperature, with a similar
behavior for magnetization. At the critical temperature, $T=T_c$, the parameter $\rho$ manifests an unforeseen discontinuous 
transition; nonetheless, the critical temperature approximates the expected value. Here, we present the graphical representation of 
the magnetization and renormalization parameter results for \ce{RbMnF_3}. The solid diamonds in the plot represent the elastic neutron scattering 
data, which have been extracted from Ref. \cite{jap111.07E145}.}
\label{fig.magnetization}
\end{figure}

For temperatures $T < T_c$, the system exhibits spontaneous symmetry breaking, and the magnetization develops long-range order along a particular direction in the $xy$-plane, 
defined by the angle $\phi$, such that the transverse component of the magnetization is given by $\mathbf{m}_\perp=m_\perp(\cos\phi,\sin\phi,0)$, where
\begin{align}
    m_\perp\approx \tilde{S}\left[1-\frac{\langle(S^z)^2\rangle_0}{2\tilde{S}^2}\right]e^{-\langle\varphi^2\rangle_0/2}.
\end{align}
Fig. (\ref{fig.magnetization}) illustrates the temperature dependence of the renormalization parameter $\rho$ and the planar magnetization $m_\perp$
for the compound \ce{RbMnF3}, which will be elaborated upon in the following paragraphs. The renormalization parameter diminishes with increasing temperature 
and abruptly approaches zero near the critical temperature $T_c$. In the analyses of both $\rho$ and $m_\perp$, the abrupt change near 
$T_c$ is attributed to the exponential factor, which depends inversely on $\rho$. Consequently, this results in an incorrect first-order transition for 
magnetization. Such anomalies are frequently observed in theories founded on harmonic expansions. Nonetheless, despite this limitation, for temperatures $T<T_c$, the QSCHA 
method yields highly accurate results, with the formalism demonstrating discrepancies only at temperatures approaching $T_c$. It is worth 
noting that even within traditional spin representations, such as the HP formalism, accurately characterizing the thermodynamics close to 
the critical temperature presents challenges. Although there is excellent concordance in the low-temperature limit when employing the 
Linear Spin-Wave (LSW) approximation, the HP formalism yields poor results at the critical temperature $T_c$. In such instances, the 
inclusion of quartic-order terms is essential, as these terms renormalize the spin-wave energy, thus providing a more plausible estimation 
of the critical temperature \cite{pps82.992}. Conversely, the critical temperature $T_c$ derived from the harmonic approximation is in close 
alignment with the actual critical temperature. Hence, despite the less accurate behavior for temperatures approaching $T_c$ from below, 
we shall regard $T_c$ as a sufficiently precise estimation of the critical temperature.

\begin{figure}[ht]
\includegraphics[width=\linewidth]{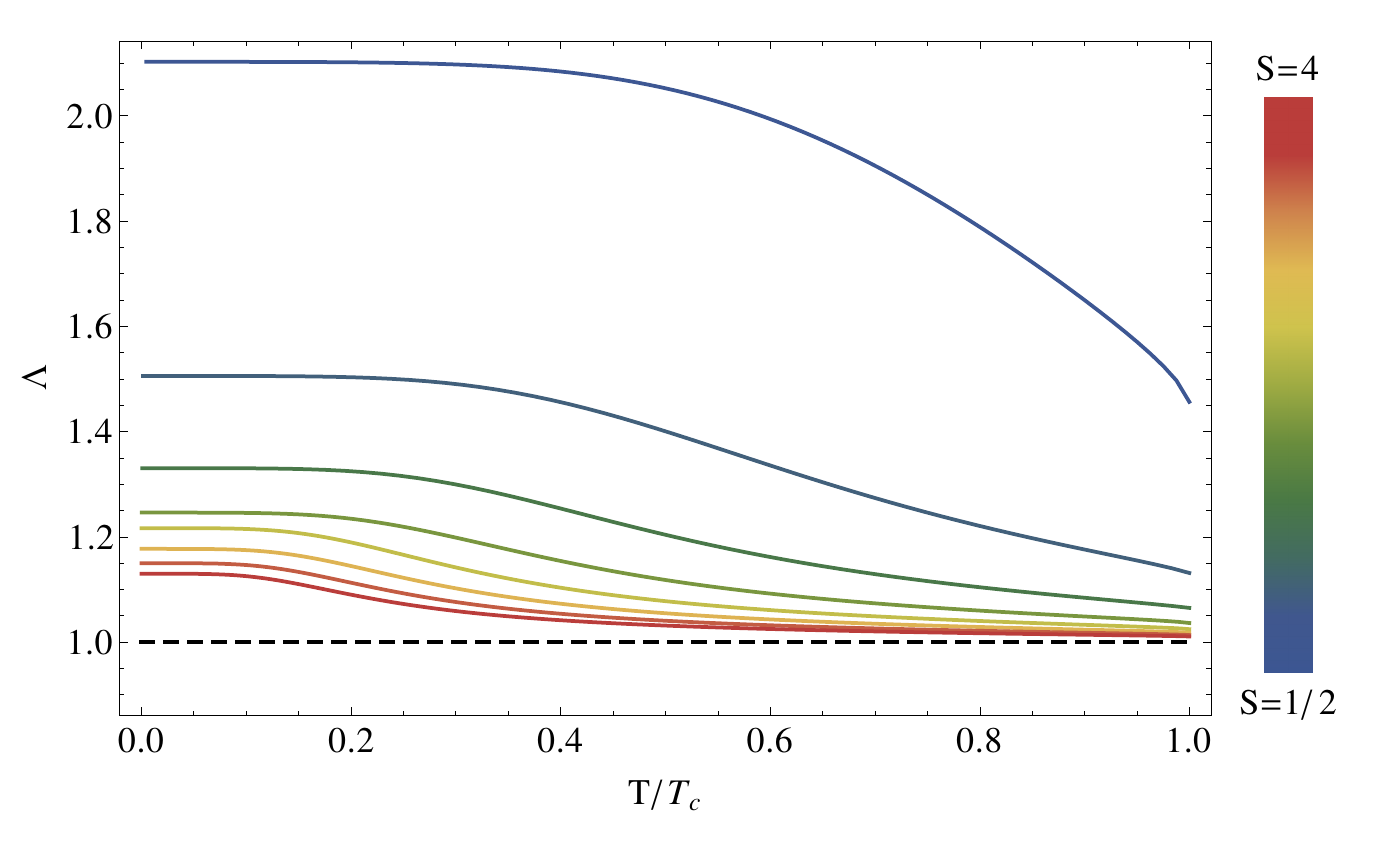}
\caption{The temperature dependence of the quantum correction factor is analyzed for spin values ranging from $S = 1/2$ to $S = 4$, with an incremental spacing of $\Delta S=1/2$. 
It is observed that the effect is more pronounced for smaller spin values, and the quantum effects disappear in the classical limit for 
which $S\gg 1$ (represented by the dashed line). The curves are determined from the XY model; however, other models exhibit analogous behavior.}
\label{fig.Lambda}
\end{figure}

The investigation of the quantum correction factor reveals its dependency on spin magnitude. Fig. (\ref{fig.Lambda}) illustrates the 
$\Lambda$ function for various spin values across the entire temperature range $0\leq T\leq T_c$, considering $\lambda=0$ (a similar 
behavior is observed for $\lambda>0$). Evidently, the correction effect is more pronounced for $S=1/2$, whereas $\Lambda$ converges toward 
unity in the classical regime ($S\gg 1$). This observed behavior manifests in the differential thermodynamic analysis pertaining to 
small versus large spin magnitudes, as will be elucidated in the next results. Moreover, the negligible influences observed in 
systems with large spin values justify the considerable efficacy of the conventional SCHA in describing spin models where the spin quantum 
number $S$ exceeds 2, as reported in the literature \cite{jmmm562.169778,jmmm587.171267}. In the particular case of the XY model, it 
is feasible to evaluate the quantum correction at $T=0$. At zero temperature, the $u_q$ and $v_q$ functions are expressed as 
$u_q = v_q = \epsilon_q/2$, thereby simplifying the quantum correction to $\Lambda = e^{\Xi}$. Consequently, the self-consistent equation reduces 
to the expression $\rho(0) = (1 - \langle(S^z)^2\rangle_0/\tilde{S}^2)$. It should be noted that, within this context, the renormalization 
parameter exclusively considers quantum fluctuations arising from the Heisenberg uncertainty principle between spin components.
The expectation value of $(S^z)^2$ is derived using the findings presented in Appendix (\ref{appendix}), and is expressed as 
\begin{align}
    \langle(S^z)^2\rangle_0 = \frac{1}{N}\sum_q\langle\bar{S}_q^zS_q^z\rangle_0 \overset{\text{(T=0)}}{=}\frac{\sqrt{\rho(0)}\tilde{S}}{2},
\end{align}
where we employ the approximation $\sum_q \sqrt{1-\gamma_q}/N=0.9747\ldots\approx 1$. By solving the equation for $\rho(0)$, 
the solution is found to be
\begin{align}
    \rho(0) = \left[\sqrt{1+(4\tilde{S})^{-1}}-(4\tilde{S})^{-1} \right]^2.
\end{align}
By following a similar procedure, it is determined that $\Xi = (2\tilde{S}\sqrt{\rho(0)})^{-1}$ at zero temperature results in
\begin{align}
    \label{eq.Lambda}
    \Lambda(T=0) = \exp\left(\frac{2}{\sqrt{16S(S+1)+1}-1} \right).
\end{align}
For $S = 1/2$, the quantum correction reaches its maximum with $\Lambda(0) \approx 2.15$, whereas $\Lambda(T)$ approaches unity as 
$S$ significantly exceeds 1, in accordance with theoretical expectations. 

\begin{table}[ht]
\begin{ruledtabular}
\begin{tabular}{cccccc}
Model & Lattice & MC & LSW & SCHA & QSCHA\\
\colrule
 & SC & 2.02 & 4.22 & 1.29 & 2.03 \\
FM XY\footnote[1]{Can. J. Phys. 50, 129 (1972)} & BCC & 2.90 & 5.96 & 1.72 & 2.71\\
 & FCC & 4.52 & 9.14 & 2.57 & 4.08 \\
\colrule
 & SC & 1.68 & 3.41 & 0.99 & 1.71\\
FM Heisenberg\footnote[2]{Phys. Rev. B 107, 235151 (2023)} & BCC & 2.52 & 5.01 & 1.40 & 2.38 \\
 & FCC & 4.01 & 7.83 & 2.17 & 3.65\\
 \colrule
AFM Heisenberg\footnote[3]{Phys. Rev. Lett. 80, 5196 (1998)} & SC & 0.95 & 3.39 & 0.63 & 0.95\\
\end{tabular}
\end{ruledtabular}
\caption{A comparative examination of the critical temperatures (in units of $J/k_B$) associated with various three-dimensional 
models is conducted. Herein, MC refers to temperatures derived via MC simulation, whereas LSW denotes results obtained through the Linear 
Spin-wave approximation within the Holstein-Primakoff formalism. The last two columns show a comparison between the standard SCHA 
and the QSCHA results. All models are defined by the spin $S=1/2$.}
\label{tab.MC}
\end{table}

To verify the efficacy of the QSCHA, we employ this formalism to examine the critical temperature transitions across various 
well-documented magnetic models. In Table (\ref{tab.MC}), we present the data pertaining to the $S=1/2$ XY and Heisenberg models, 
derived from MC simulations, LSW analysis via the HP formalism, the conventional SCHA, and the novel QSCHA. The HP 
results were ascertained by identifying the temperature at which the transverse magnetization $m_\perp=S-N^{-1}\sum_q n_q$ becomes null, 
whereas the SCHA results are interpreted as the temperature at which a discontinuous change in $\rho$ occurs. The QSCHA yields superior 
results, whereas the LSW theory predicts critical temperatures that are approximately twice as high as the anticipated values. 
As previously discussed, neglecting magnon interactions near the critical temperature is not a scientifically valid approach, resulting 
in the poor results obtained. Superior outcomes are achieved when quartic or higher-order terms are incorporated; nevertheless, 
the implementation of interactions within the HP formalism exhibits greater complexity compared to the QSCHA framework.

\begin{table}[ht]
\begin{ruledtabular}
\begin{tabular}{cccccc}
Compound& Spin & Exp. & LSW & SCHA & QSCHA\\
\colrule
\ce{KMnF_3}\footnote[4]{J. Phys. Colloques 32, 1184 (1971)} & 5/2 & 88 K & 199.1 K & 92.4 K & 85.1 K\\
\ce{RbMnF_3}\footnote[5]{Proc. Phys. Soc. 87, 501 (1966)} & 5/2 & 83 K & 221.9 K & 83.9 K & 76.8 K\\
\end{tabular}
\end{ruledtabular}
\caption{A comparative analysis of the critical temperatures for three-dimensional magnetic models. 
The critical temperature $T_c$ has been determined via empirical measurements. All compounds exhibit a SC lattice structure. 
The abbreviation LSW refers to Linear Spin-wave approximation, implemented within the formalism of 
the Holstein-Primakoff representation. The final two columns present a comparative analysis between the conventional SCHA and 
the QSCHA results.}
\label{tab.exp}
\end{table}

In Table (\ref{tab.exp}), we present a comparative analysis of the theoretical predictions derived from LSW theory, the 
conventional SCHA, and the QSCHA against empirical measurements of the critical temperature for three magnetic materials exhibiting an 
SC lattice structure. In this Table, the compounds \ce{KMnF_3} and \ce{RbMnF_3} are classified as nearly isotropic AFM materials. 
The determination of critical temperatures was conducted following the previously described procedure. For optimal accuracy in the results, the exchange 
coupling constant $J$ was computed via spectrum curve analysis. Given the linear dependency of energy on $J$, the least-squares fitting method 
was employed to determine the value of $J$. Specifically, within the LSW framework, the critical temperature $T_c$ was determined utilizing the constant 
derived from the equation:
\begin{align}
J=\frac{1}{2z S}\frac{\displaystyle\sum_i \epsilon_i\sqrt{(1-\gamma_{q_i})(1\pm\gamma_{q_i})}}{\displaystyle\sum_i (1-\gamma_{q_i})(1\pm\gamma_{q_i})},    
\end{align}
where $\epsilon_i$ denotes the energies corresponding to the wave-vector $\mathbf{q}_i$, and the sum is performed over the 
experimental dataset. 

For the SCHA approaches, the procedure exhibits a slight deviation from the LSW scenario. We introduce the temperature-dependent 
renormalized coupling, denoted as $J_r(T)=J\sqrt{\rho(T)}$. We recognize that $J_r$ signifies the exchange coupling determined 
from empirical observations under finite temperature conditions, whereas the bare value $J$ is considered the intrinsic parameter 
exclusively when $\rho=1$. Consequently, the application of the least squares method yields the following expression:
\begin{align}
J_r = \frac{1}{2z\tilde{S}}\frac{\displaystyle\sum_i \epsilon_i\sqrt{(1-\gamma_{q_i})(1\pm\gamma_{q_i})}}{\displaystyle\sum_i (1-\gamma_{q_i})(1\pm\gamma_{q_i})}. 
\end{align}
Upon the establishment of $J_r$, we subsequently employ the self-consistent equations to determine the renormalization 
parameter $\rho$ at the same temperature as that of the energy spectrum experiment, thereby allowing for the determination 
of the bare constant as $J = J_r/\sqrt{\rho}$. 

\begin{figure}[ht]
\includegraphics[width=0.9\linewidth]{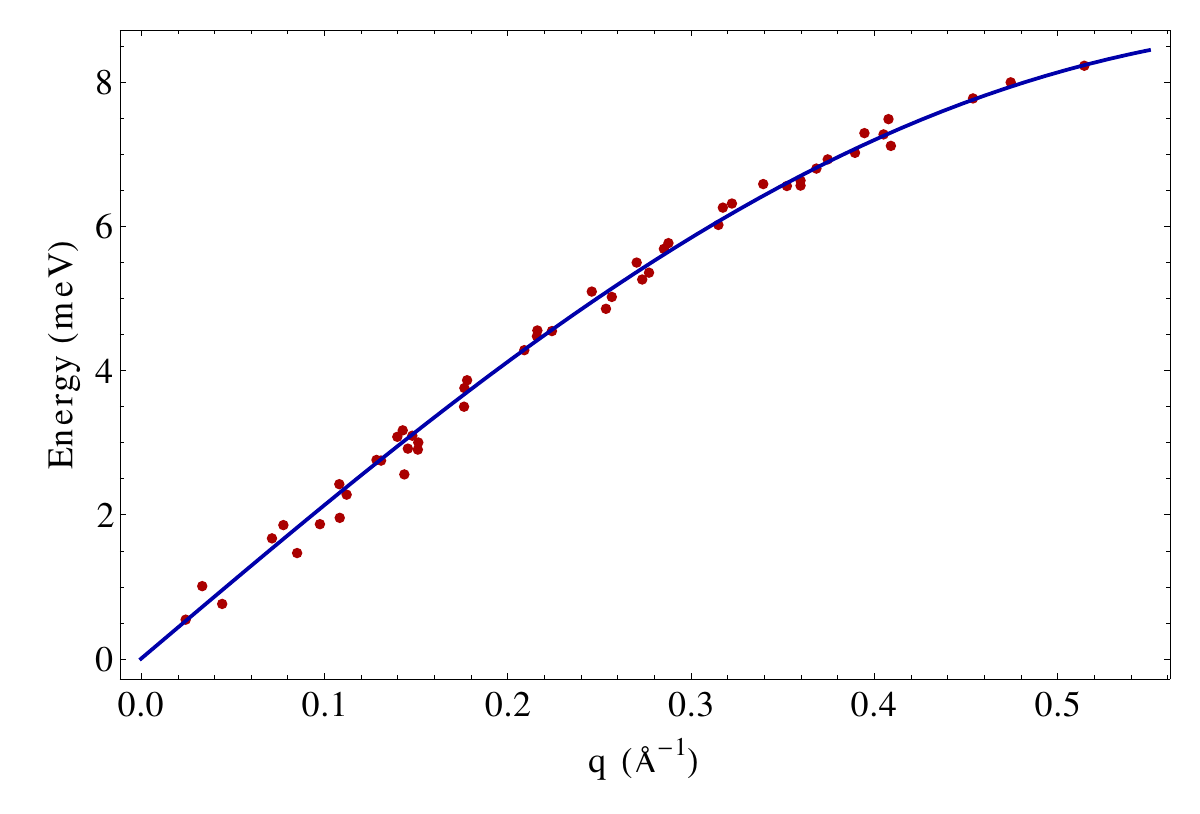}
\caption{The theoretical results obtained by applying the QSCHA to the spectral energy analysis of the \ce{RbMnF_3} compound at $T = 4.2$ K, with wave vectors probed along the (110) crystallographic plane. The experimental data have been extract from Ref. \cite{pps87.501}.}
\label{fig.spectrum}
\end{figure}

To facilitate comparison with empirical observations, the QSCHA is employed for the compound Rubidium manganese(III) 
fluoride (\ce{RbMnF3}), which adopts an ideally cubic perovskite structure with a lattice parameter of $a = 4.24~\text{\r{A}}$. 
Each \ce{Mn^{3+}} ion occupies the center of an octahedron formed by six \ce{F^-} ions, and the \ce{Mn^{3+}} ions interact via 
antiferromagnetic superexchange through these fluoride ions. The primary magnetic interactions occur between each \ce{Mn^{3+}} ion
and its six nearest-neighbor \ce{Mn^{3+}} ions in the lattice. Fig. (\ref{fig.spectrum}) presents the experimental data for the 
spectral energy of \ce{RbMnF_3}, together with the best fit obtained within the QSCHA framework. By applying the least-squares 
method to fit the spectrum curve at $T=4.2$ K, we obtain a temperature-independent coupling constant of $J = 3.10$ K. This 
value is then used to calculate the full temperature dependence of both the renormalization parameter and the magnetization, as 
illustrated in Fig. (\ref{fig.magnetization}). 

\begin{figure}[ht]
\includegraphics[width=0.9\linewidth]{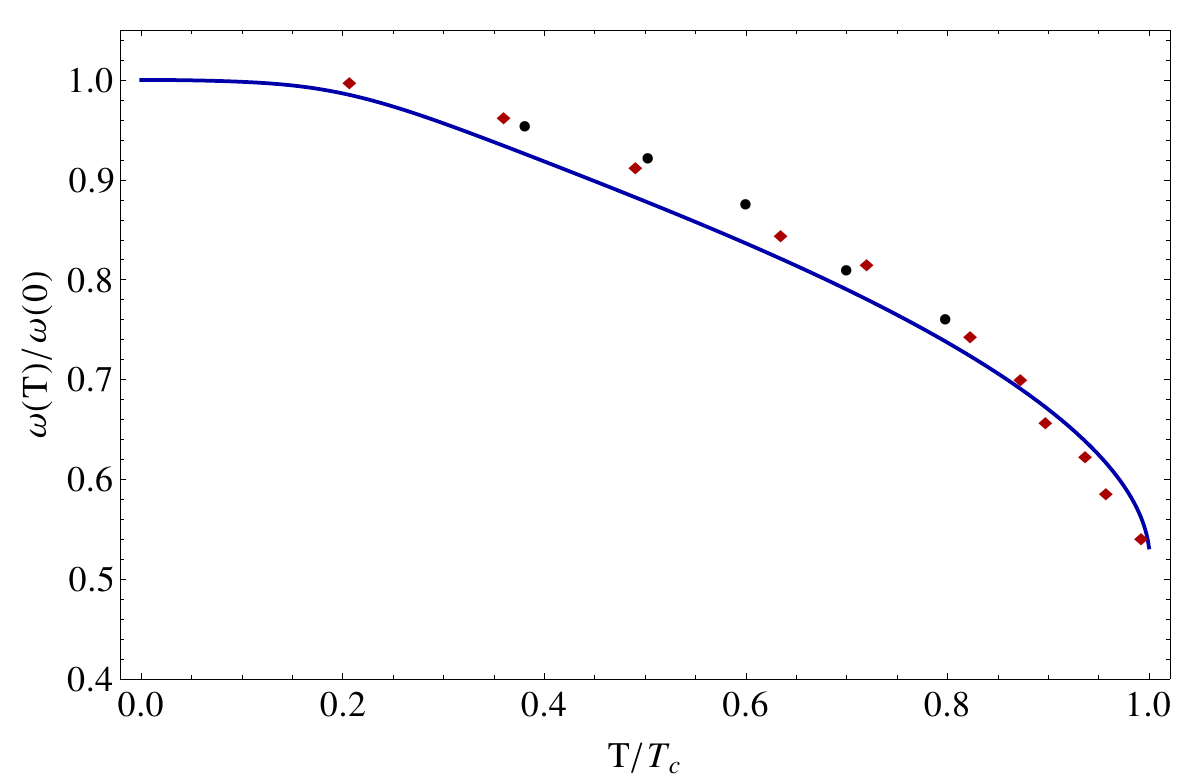}
\caption{Renormalized energy of boundary-zone magnons. The solid line represents the results obtained from the QSCHA, the 
black circles correspond to neutron scattering data from Ref. \cite{nis72}, and the red diamonds indicate Raman scattering 
data from Ref. \cite{ssc25.241}.}
\label{fig.omega}
\end{figure}

Interestingly, and in contrast to the other cases, the quantum SCHA leads to a slightly less accurate agreement with the reference 
data than the conventional approach. Nevertheless, the renormalization parameter still produces reliable results for temperatures in 
the vicinity of the transition point.In the vicinity of $T = 0$ K, the magnetization exhibits a power-law decrease proportional to 
$T^{2.03}$, in agreement with conventional spin-wave theory, which predicts a $T^2$ dependence. However, an analysis over the extended 
temperature range $0 \leq T \leq 0.85\,T_c$ yields $m_\perp \approx 1 - 0.43\,(T/T_c)^{2.32}$, which is consistent with 
the power-law dependence $T^{2.38}$ extracted from experimental magnetization data \cite{jap111.07E145}. Additionally, we present a comparison between the temperature dependence of the renormalized energy obtained from the QSCHA formalism and experimental data from two-magnon scattering, which exhibit a reduction in excitation energy, as illustrated in Fig. (\ref{fig.omega}). As can be observed, there is a good quantitative agreement between the theoretical zone-boundary magnon frequency at temperature $T$, denoted by $\omega(T)$, and the corresponding experimental measurements \cite{nis72,ssc25.241}.

The results obtained from \ce{RbMnF_3} are comparable to those observed in the other compounds. Analyzing the experimental data, 
it is evident that the outcomes predicted by LSW theory exhibit a large variance. Despite this, the critical temperature obtained through 
both the conventional SCHA and the QSCHA remains comparable. For the AFM materials, the spin value is considerable, decreasing the 
role of the quantum correction given by $\Lambda$.

\section{Summary and conclusion}
\label{sec.conclusion}
The SCHA offers a straightforward formalism for the examination of magnetic models. Despite the inherently quadratic structure of the Hamiltonian, 
the SCHA incorporates fluctuation corrections via a renormalization parameter dependent on temperature, which is determined through a self-consistent equation.
Consequently, this model can describe thermodynamic properties more accurately, even in proximity to the critical temperature, compared 
to the LSW approach. The precision of results derived from the SCHA is notably high for models with large spin values, although it remains qualitative for 
systems where $S=1/2$. In scenarios characterized by small spin , minor adjustments, such as substituting $S$ with $\sqrt{S(S+1)}$, 
have been implemented in the SCHA to enhance the accuracy of the results, although these modifications do not always yield satisfactory accuracy.

In this study, we present a detailed development of the QSCHA. In contrast to other studies in the literature, we do not presuppose the semiclassical 
limit throughout our analysis, yielding more precise results, particularly for quantum models with $S=1/2$. Through the application of an 
appropriate demonstration, we have derived a self-consistent equation analogous to those found in traditional SCHA; however, our findings reveal the 
presence of an additional multiplicative factor, denoted as $\Lambda(T)$. The term $\Lambda(T)$ is an effect exclusively of quantum origin, 
which does not manifest within the semiclassical framework. Indeed, at zero temperature, we find that $\Lambda(0)$ is given by Eq. (\ref{eq.Lambda}),
which yields a significant correction for $S=1/2$, while it tends to unity for large spin values. Furthermore, additional subsidiary 
explanations are provided, such as the necessary substitution of $S$ with $\tilde{S}=\sqrt{S(S+1)}$. 

To verify the developed method, we compare the results from QSCHA with MC simulations and experimental data obtained from the
literature. The comparison shows a substantial improvement for magnetic quantum models with $S=1/2$, while the difference between the conventional
SCHA and the new QSCHA is minor for models with $S\geq 2$, as expected. 

In conclusion, the QSCHA has proven to be a superior alternative to the conventional LSW formalism, especially near the critical temperature. 
This is attributed to its straightforward quadratic Hamiltonian structure while simultaneously incorporating quantum and thermal 
corrections that account for fluctuations arising from higher-order terms typically neglected in the harmonic approximation. 
Moreover, owing to its formulation based on canonically conjugate operators, the standard SCHA has been effectively applied in spintronic 
applications involving resonance phenomena. Consequently, the novel QSCHA emerges as a promising formalism for exploring advanced 
quantum technologies.

This study was financed in part by the Coordenação de Aperfeiçoamento de Pessoal de Nível Superior – Brasil 
(CAPES) - Finance Code 001, and by the National Council for Scientific and Technological Development – CNPq. 

\appendix
\section{Non-interacting field averages} 
\label{appendix}
Performing the Fourier transform as defined by
\begin{align}
    \varphi_i(\tau)=\frac{1}{\beta\hbar\sqrt{N}}\sum_{q}\sum_{\omega_n}\varphi_{kn}e^{i(\mathbf{q}\cdot\mathbf{r}_i-\omega_n\tau)},
\end{align}
and similar transform for $S_i^z(\tau)$, the action of the non-interacting model is written as
\begin{align}
    \mathcal{A}_0=&\frac{1}{2\beta\hbar}\sum_{q,\omega_n}(-\hbar\omega_n\bar{S}_{qn}^z\varphi_{qn}+\hbar\omega_n S_{qn}^z\bar\varphi_{qn}+\nonumber\\
    &+h_q^\varphi\bar{\varphi}_{qn}\varphi_{qn}+h_q^z\bar{S}_{qn}^zS_{qn}^z),
\end{align}
where $\omega_n=2\pi n/\beta\hbar$, $n\in\mathbb{Z}$, are the bosonic Matsubara frequencies. 
To eliminate the linear contributions, we define the deviation angle $\delta\varphi_{qn}=\varphi_{qn}-\phi_{qn}$, where $\phi_{qn}$ is
determined by the minimum condition:
\begin{align}
    \left.\frac{\partial\mathcal{A}_0}{\partial\varphi_{qn}}\right|_{\varphi_{qn}=\phi_{qn}}=0.
\end{align}
Therefore, in the context of the field $\delta\varphi_{qn}$, the action is given by
\begin{align}
    \mathcal{A}_0&=\frac{1}{2\beta\hbar}\sum_q\sum_{\omega_n}\left[\frac{\hbar^2(\omega_q^2+\omega_n^2)}{h_q^\varphi}\bar{S}_{qn}^zS_{qn}^z+\right.\nonumber\\
    &\left.+h_q^\varphi\delta\bar{\varphi}_{qn}\delta\varphi_{qn}\right],
\end{align}
wherein the partition function can be decomposed as $Z_0=Z_0^\varphi Z_0^z$. Specifically, $Z_0^\varphi$ is defined by
\begin{align}
    Z_0^\varphi&=\prod_{q,n}\int \dd(\delta\bar{\varphi}_{qn})\dd(\delta\varphi_{qn})e^{-(h_q^\varphi/\beta\hbar^2)\bar{\delta\varphi}_{qn}\delta\varphi_{qn}}\nonumber\\
    &=(\det \beta h_q^\varphi)^{-1/2},
\end{align}
where we disregard constant factors that do not affect the dynamics and serve as a multiplicative normalization constant.
A similar procedure results in
\begin{align}
    Z_0^z=\left[\det \left(\frac{\beta\hbar^2(\omega_q^2+\omega_n^2)}{h_q^\varphi}\right)\right]^{-1/2},
\end{align}
providing $Z_0=[\det \beta^2\hbar^2(\omega_q^2+\omega_n^2)]^{-1/2}$. From the partition function, we derive 
$\langle \bar{S}_{qn}^z S_{qn}^z\rangle_0=\beta h_q^\varphi(\omega_q^2+\omega_n^2)^{-1}$ and the Matsubara sum yields
\begin{align}
    \langle \bar{S}_q^z(\tau)S_q^z(0)\rangle_0&=\frac{h_q^\varphi}{\beta\hbar}\sum_{\omega_n}\frac{e^{i\omega_n \tau}}{\omega_n^2+\omega_q^2}\nonumber\\ 
    &=\frac{h_q^\varphi}{2\epsilon_q}\frac{\cosh(\beta\epsilon_q/2+\omega_q\tau)}{\sinh(\beta\epsilon_q/2)}.
\end{align}
The application of an analogous approach to the angle terms yields $\langle \bar{\varphi}_{qn} \varphi_{qn}\rangle_0=\beta h_q^z(\omega_q^2+\omega_n^2)^{-1}$, and
\begin{align}
    \langle \bar{\varphi}_q(\tau)\varphi_q(0)\rangle_0=\frac{h_q^z}{2\epsilon_q}\frac{\cosh(\beta\epsilon_q/2+\omega_q\tau)}{\sinh(\beta\epsilon_q/2)}.
\end{align}
For equal time, we recover the expected results $\langle \bar{S}_q^zS_q^z\rangle_0=(h_q^\varphi/2\epsilon_q)\coth(\beta\epsilon_q/2)$ and
$\langle \bar{\varphi}_q \varphi_q \rangle_0=(h_q^z/2\epsilon_q)\coth(\beta\epsilon_q/2)$, alongside the expression for the expected value
$\langle\hat{H}_0\rangle_0=\sum_q\epsilon_q(n_q+1/2)$.

\bibliographystyle{apsrev4-2}
\bibliography{bibliography.bib}
\end{document}